\documentclass[sigconf]{acmart}
\usepackage{amsmath,amsfonts}
\usepackage{algorithmic}
\usepackage{graphicx}
\usepackage{textcomp}
\usepackage{xcolor}
\usepackage{url}
\usepackage{multirow}
\usepackage{tabularx}
\usepackage{subcaption}
\usepackage{float}
\usepackage{rotating}

\usepackage{array, tabularx}
\newcolumntype{Y}{>{\centering\arraybackslash}X}

\usepackage{booktabs}

\usepackage{cellspace}

\AtBeginDocument{%
  \providecommand\BibTeX{{%
    Bib\TeX}}}

\setcopyright{acmlicensed}
\copyrightyear{2025}
\acmYear{2025}
\acmDOI{XXXXXXX.XXXXXXX}
\acmISBN{978-1-4503-XXXX-X/2018/06}

\def\BibTeX{{\rm B\kern-.05em{\sc i\kern-.025em b}\kern-.08em
    T\kern-.1667em\lower.7ex\hbox{E}\kern-.125emX}}
\begin{document}

\title{An Empirical Study of LLM-Based Code Clone Detection}

\author{Wenqing Zhu}
\email{zhuwqing1995@ertl.jp}
\authornotemark[1]
\affiliation{%
  \institution{Nagoya University}
  \city{Nagoya}
  \country{Japan}
}

\author{Norihiro Yoshida}
\email{norihiro@fc.ritsumei.ac.jp}
\affiliation{%
  \institution{Ritsumeikan University}
    \city{Osaka}
  \country{Japan}
}

\author{Eunjong Choi}
\email{echoi@kit.ac.jp}
\affiliation{%
  \institution{Kyoto Institute of Technology}
    \city{Kyoto}
  \country{Japan}
}
\author{Yutaka Matsubara}
\email{yutaka@ertl.jp}
\affiliation{%
  \institution{Nagoya University}
    \city{Nagoya}
  \country{Japan}
}
\author{Hiroaki Takada}
\email{hiro@ertl.jp}
\affiliation{%
  \institution{Nagoya University}
    \city{Nagoya}
  \country{Japan}
}
\renewcommand{\shortauthors}{Zhu et al.}

\begin{abstract}

Large language models (LLMs) have demonstrated remarkable capabilities in various software engineering tasks, such as code generation and debugging, because of their ability to translate between programming languages and natural languages.
Existing studies have demonstrated the effectiveness of LLMs in code clone detection. However, two crucial issues remain unaddressed: the ability of LLMs to achieve comparable performance across different datasets and the consistency of LLMs' responses in code clone detection.
To address these issues, we constructed seven code clone datasets and then evaluated five LLMs in four existing prompts with these datasets. The datasets were created by sampling code pairs using their Levenshtein ratio from two different code collections, CodeNet and BigCloneBench.
Our evaluation revealed that although LLMs perform well in CodeNet-related datasets, with o3-mini achieving a 0.943 F1 score, their performance significantly decreased in BigCloneBench-related datasets.
Most models achieved a high response consistency, with over 90\% of judgments remaining consistent across all five submissions. The fluctuations of the F1 score affected by inconsistency are also tiny; their variations are less than 0.03.
\end{abstract}




\maketitle

\section{Introduction}

Code clones are pairs of identical or similar code fragments. These clones are generally created when developers reuse functionality by copying and pasting existing source code \cite{roy2007survey, roy2009comparison,al2005cloning,baker1993clone}.
They can be classified into syntactic and semantic clones \cite{roy2007survey}.
Syntactic clones have high syntactic similarity, while semantic clones may have low syntactic similarity but identical functionality. 
Detecting code clones has traditionally been treated as a classification problem \cite{NICAD,SourcererCC,ccfinder}. 
Traditional code clone detectors generally employ threshold-based similarity measurements to identify code fragments as code clones. 
While these detectors effectively identify syntactic clones, they struggle to detect semantic clones because the same functionality can be implemented using different syntax.

In recent years, the Large Language Model (LLM) has demonstrated notable performance across various software engineering tasks, including code generation, debugging, and comprehension \cite{LLMAPR-xia2023conversational, llmCodeGenerationNonDeterminism-ouyang2023,LLMDebug-tian2024, LLMCodeUnderstandingICSE24-nam2024,LLMAPR_Higo,LLMCodeReview_rasheed2024ai,LLMTestingSurvey}. 
LLM has the potential for accurate code clone detection as it can capture 
deeper relationships between code clones, even when their syntax differs. 
LLM-based code clone detectors allow developers to identify code clones using zero-shot or few-shot prompting. In this process, developers provide a prompt  that includes a pair of code fragments and ask the model to evaluate their similarity. 
The model then generates an output, which is interpreted as a Boolean value either directly or with the help of a discriminator to determine whether the code fragments are clones.

Previous studies have demonstrated the effectiveness of LLMs in code clone detection \cite{LLMCCD_ARXIV2023, LLMCCD_ICPC2024}.
Khajezade et al. demonstrated the effectiveness of \textit{GPT-3.5-turbo} with their proposed prompts in detecting code clones, particularly for \texttt{Java}-\texttt{Java} and \texttt{Java}-\texttt{Ruby} clone pairs \cite{LLMCCD_ICPC2024}.
Dou et al. evaluated the performance of LLMs with prompt engineering techniques and found that \textit{GPT-3.5-turbo} and \textit{GPT-4} outperform traditional code clone detectors in identifying semantic code clones \cite{LLMCCD_ARXIV2023}. 
While these studies highlight the potential of LLMs for code clone detection, a deeper understanding of their capabilities in detecting code clones is necessary. In particular, evaluating their generalization ability and response consistency is crucial to ensure the practical application.
\noindent

\begin{figure*}[]
\centering
    \graphicspath{{./img/}}
        \includegraphics[width=\textwidth]{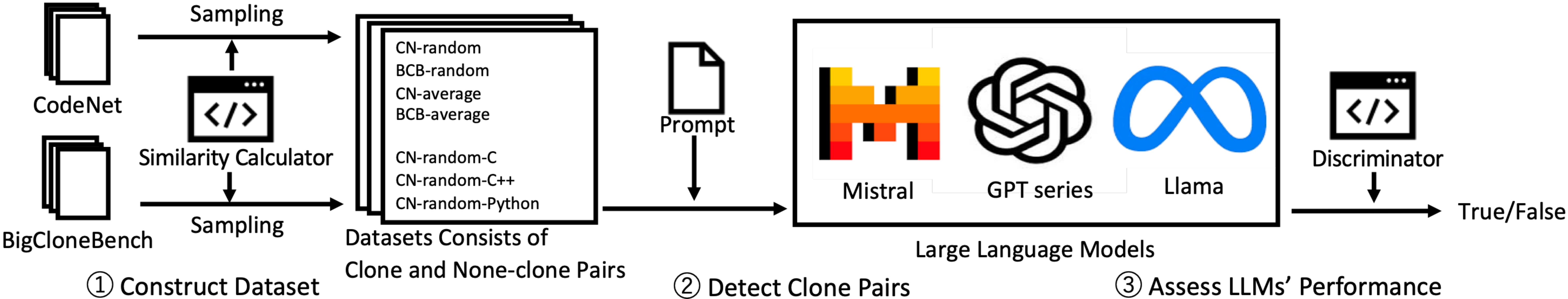}
    \caption{Experiment Process}
    \label{fig:ExperimentProcess}
\end{figure*}
 
Regarding generalization, while Deep Learning (DL)-based clone detectors perform well in detecting semantic clones \cite{CCLearner, ASTNN, CodeBERTfeng2020codebertpretrainedmodelprogramming}, they struggle to generalize across diverse datasets \cite{ChoiAIDetectorGeneralizability}. 
Since LLMs are trained on significantly larger datasets than DL models, they may offer better generalization. However, this potential still requires thorough investigation.
Similarly, response consistency is a critical factor. LLMs are designed for contextual generation, generating varied outputs rather than identical responses to the same input. However, for code clone detection, judgments for each candidate must remain stable and deterministic. Therefore, investigating the response consistency of LLM in detecting code clones is necessary for reliable detection.
Despite their importance, these issues have not been thoroughly investigated.

To address these issues, this study investigates LLM-based code clone detectors to answer the following two Research Questions (RQs):

\begin{description}

\item [\textbf{RQ1:}] How accurately does each LLM detect code
clones across different datasets?

 \item [\textbf{RQ2:}] How consistent are the responses from each LLM when presented with identical input?
\end{description}

To answer these RQs, we constructed seven datasets in various programming languages (four in \texttt{Java} and one each in \texttt{Python}, \texttt{C}, and \texttt{C++}) by selecting clone and non-clone pairs from two commonly used code clone evaluation datasets \textit{CodeNet} (\textit{CN}) \cite{CodeNet} and \textit{BigCloneBench} (\textit{BCB}) \cite{BigCloneBench}. In these datasets datasets, all code pairs were labeled according to their syntactic similarity. For  \texttt{Java}, the datasets were divided into two types based on similarity distribution to ensure a fair comparison of performance across datasets.
Furthermore, we selected five LLMs (i.e., \textit{o3-mini}, \textit{GPT-4o}, \textit{GPT-4o-mini}, \textit{Llama 3.1}, and \textit{Mistral}) and evaluated their performance in code clone detection. The prompts were chosen based on existing studies \cite{LLMCCD_ARXIV2023,LLMCCD_ICPC2024}. The evaluation metrics included recall, precision, F1 score, and response consistency rate. Each model-prompt combination was evaluated in different temperature settings (a parameter to control the randomness of LLM's output) and attempt times.
The evaluation results show that LLMs demonstrated superior performance on \textit{CN}-derived datasets than on \textit{BCB}, and most models exhibited high response consistency with minimal performance variations.

The contributions of this study are listed below:
\begin{itemize}
    \item All models demonstrated significantly higher performance on datasets derived from \textit{CN}, with certain model-prompt combinations achieving F1 scores above 0.9. However, on datasets derived from \textit{BCB}, most model--prompt pairs could not accurately detect clones effectively in low-similarity ranges
    \item All LLMs, except \textit{Llama 3.1}, exhibited high response consistency, with response inconsistencies having minimal impact on F1 scores across models. Additionally, prompt selection typically affected response consistency more than the temperature settings of the LLMs.
    \item The temperature value of LLM has a small influence on code clone detection on both response consistency and F1 score.
    \item For the replicability of this study, we have published seven datasets covering four languages\footnote{All our datasets, programs, and results can be accessed by: \url{https://zenodo.org/records/15019503}}. Besides, our scripts for accessing LLMs for clone detection and analyzing the results are also included.
\end{itemize}

\section{Related Work}
\label{sec:relatedStudy}

Traditionally, code clone detection has been treated as a classification problem. 
Traditional code clone detectors typically transform the input code into intermediate representations and compare the syntactic similarity based on a predefined threshold to identify potential clones \cite{NICAD,SourcererCC,MSCCD,nasehi2007source,koschke2006clone,graph-basedCCD-CCSharp}. They effectively detect syntactic clones but struggle with semantic clones because code fragments that implement the same functionality may exhibit significant syntactic variation \cite{SourcererCC, BigCloneBench,MSCCD_JSS,wu2020scdetector,yu2019neural}.

To evaluate code clone detectors, several code clone datasets have been developed to measure the accuracy and performance of code clone detectors \cite{semanticclonebench},\cite{CodeNet},\cite{BigCloneBench},\cite{MSCCD_JSS},\cite{zhao18fse}. 
One such dataset, \textit{BCB}, is constructed from commonly implemented \texttt{Java} functions mined from open-source software. It also provides similarity data for clone pairs \cite{BigCloneBench}. 
Other datasets are constructed using correct code submissions for identical problems from programming competitions. For example, Google Code Jam\cite{zhao18fse} provides \texttt{Java} submissions, and \textit{CN} \cite{CodeNet} encompasses submissions across 55 different programming languages.
More recently, SemanticCloneBench was constructed by collecting semantic clones from Stack Overflow answers\cite{semanticclonebench}. Similarly, Zhu et al.\cite{MSCCD_JSS} published code clone datasets supporting four programming languages, including 14 problems from \textit{CN}. This dataset contains clone pairs with precomputed Levenshtein ratios to reduce computational time.

\begin{table*}[t]
\caption{Detail of Datasets Used in This Work}
\label{Tab:DetailsOfJavaDatasets}

\resizebox{\textwidth}{!}{
\begin{tabular}{llccccccccccccclc}
\toprule

\multirow{3}{*}[-0.8em]{Language} &\multirow{3}{*}[-0.8em]{Dataset} & \multicolumn{12}{c}{Code pairs in each similarity range}                                                                                                                                                                                                                                                  & \multicolumn{2}{c}{\multirow{2}{*}[-0.4em]{Sum}}   & \multirow{3}{*}[-0.8em]{RQ} \\ \cmidrule(lr){3-14}
                         & & \multicolumn{2}{c}{{[}0,0.2)}                      & \multicolumn{2}{c}{{[}0.2,0.4)}                    & \multicolumn{2}{c}{{[}0.4,0.6)}                    & \multicolumn{2}{c}{{[}0.6,0.8)}                    & \multicolumn{2}{c}{{[}0.8,1.0)}                    & \multicolumn{2}{c}{{[}1.0,1.0{]}} & \multicolumn{2}{c}{}                    & \\ \cmidrule(lr){3-16} 
                         & & \multicolumn{1}{c}{TC}  & \multicolumn{1}{c}{FC}  & \multicolumn{1}{c}{TC}  & \multicolumn{1}{c}{FC}  & \multicolumn{1}{c}{TC}  & \multicolumn{1}{c}{FC}  & \multicolumn{1}{c}{TC}  & \multicolumn{1}{c}{FC}  & \multicolumn{1}{c}{TC}  & \multicolumn{1}{c}{FC}  & \multicolumn{1}{c}{TC}     & FC   & \multicolumn{1}{c}{TC}       & FC    &   \\ \midrule
\multirow{5}{*}[-0.6em]{Java}    & \textit{CN-random}  & \multicolumn{1}{c}{140} & \multicolumn{1}{c}{56}  & \multicolumn{1}{c}{140} & \multicolumn{1}{c}{399} & \multicolumn{1}{c}{140} & \multicolumn{1}{c}{242} & \multicolumn{1}{c}{140} & \multicolumn{1}{c}{3}   & \multicolumn{1}{c}{140} & \multicolumn{1}{c}{0}   & \multicolumn{1}{c}{-}      & -    & \multicolumn{1}{c}{700}      & 700  & RQ1,2    \\ \cmidrule(lr){2-17}
               &\textit{BCB-random}  &\multicolumn{1}{c}{140} & \multicolumn{1}{c}{56}  & \multicolumn{1}{c}{140} & \multicolumn{1}{c}{399} & \multicolumn{1}{c}{140} & \multicolumn{1}{c}{242} & \multicolumn{1}{c}{140} & \multicolumn{1}{c}{3}   & \multicolumn{1}{c}{140} & \multicolumn{1}{c}{0}   & \multicolumn{1}{c}{-}      & -    & \multicolumn{1}{c}{700}      & 700   & RQ1   \\ \cmidrule(lr){2-17}
              &\textit{CN-average}  & \multicolumn{1}{c}{140} & \multicolumn{1}{c}{140} & \multicolumn{1}{c}{140} & \multicolumn{1}{c}{140} & \multicolumn{1}{c}{140} & \multicolumn{1}{c}{140} & \multicolumn{1}{c}{140} & \multicolumn{1}{c}{0}   & \multicolumn{1}{c}{140} & \multicolumn{1}{c}{0}   & \multicolumn{1}{c}{140}    & 0    & \multicolumn{1}{c}{840}      & 420  & RQ1    \\ \cmidrule(lr){2-17}
               &\textit{BCB-average} & \multicolumn{1}{c}{140} & \multicolumn{1}{c}{140} & \multicolumn{1}{c}{140} & \multicolumn{1}{c}{140} & \multicolumn{1}{c}{140} & \multicolumn{1}{c}{140} & \multicolumn{1}{c}{140} & \multicolumn{1}{c}{140} & \multicolumn{1}{c}{118} & \multicolumn{1}{c}{118} & \multicolumn{1}{c}{140}    & 0    & \multicolumn{1}{c}{818}      & 678   & RQ1   \\ \midrule
C & \textit{CN-random-C}               & \multicolumn{1}{c}{140} & \multicolumn{1}{c}{78} & \multicolumn{1}{c}{140} & \multicolumn{1}{c}{368} & \multicolumn{1}{c}{140} & \multicolumn{1}{c}{251} & \multicolumn{1}{c}{140} & \multicolumn{1}{c}{3} & \multicolumn{1}{c}{118} & \multicolumn{1}{c}{0} & \multicolumn{1}{c}{-}    & -    & \multicolumn{1}{c}{700}      & 700  & RQ1    \\ \midrule
C++ & \textit{CN-random-C++}               & \multicolumn{1}{c}{140} & \multicolumn{1}{c}{85} & \multicolumn{1}{c}{140} & \multicolumn{1}{c}{453} & \multicolumn{1}{c}{140} & \multicolumn{1}{c}{159} & \multicolumn{1}{c}{140} & \multicolumn{1}{c}{3} & \multicolumn{1}{c}{118} & \multicolumn{1}{c}{0} & \multicolumn{1}{c}{-}    & -    & \multicolumn{1}{c}{700}      & 700   & RQ1   \\ \midrule
Python & \textit{CN-random-Python}               & \multicolumn{1}{c}{140} & \multicolumn{1}{c}{49} & \multicolumn{1}{c}{140} & \multicolumn{1}{c}{395} & \multicolumn{1}{c}{140} & \multicolumn{1}{c}{254} & \multicolumn{1}{c}{140} & \multicolumn{1}{c}{2} & \multicolumn{1}{c}{118} & \multicolumn{1}{c}{0} & \multicolumn{1}{c}{-}    & -    & \multicolumn{1}{c}{700}      & 700  & RQ1    \\ \bottomrule
\end{tabular}
}

 \footnotesize{ 
    \flushleft{ CN: CodeNet  BCB: BigCloneBench TC: clones FC: non-clones\noindent\\
    }
    }
\end{table*}

\begin{table*}[!]
\caption{Prompts evaluated in this study}
\label{Tab:PromptsList}
\resizebox{0.98\textwidth}{!}{
\begin{tabular}{llc}
\toprule
ID & Prompt                                                                                                                                                                                                                                                                                                                                                                                                                                & \multirow{1}{*}{ Ref.}                                                   \\ \midrule
\textit{P0}  & Do code 1 and code 2 solve identical problems with the same inputs and outputs? Answer with yes or no and no explanation.                                                                                                                                                                                                                                                                                                             & \multirow{1}{*}{\cite{LLMCCD_ICPC2024}}                   \\ \midrule
\textit{P1}  & \begin{tabular}[c]{@{}l@{}}Please analyze the following two code snippets to assess their similarity and determine if they are code clones. \\ Provide a similarity score between 0 and 10, where a higher score indicates more similarity. \\ Additionally, identify the type of code clone they represent and present a detailed reasoning process for detecting code clones. \\ The response should be `yes' or `no'.\end{tabular} & \multirow{3}{*}[-2em]{\cite{LLMCCD_ARXIV2023}} \\ \cmidrule(lr){1-2}
\textit{P2}  & \begin{tabular}[c]{@{}l@{}}Please provide a detailed reasoning process for detecting code clones in the following two code snippets. \\ Based on your analysis, respond with `yes' if the code snippets are clones or `no' if they are not.\end{tabular}                                                                                                                                                                              &                                                            \\ \cmidrule(lr){1-2}
\textit{P3}  & \begin{tabular}[c]{@{}l@{}}Please analyze the following two code snippets for code clone detection. You should first report which lines of code are more similar. \\ Then based on the report, please answer whether these two codes are a clone pair. The response should be ‘yes’ or `no'.\end{tabular}                                                                                                                       &                                                            \\ \bottomrule
\end{tabular}
}
\end{table*}

Over the past decade, DL-based code clone detectors have shown promise in detecting semantic clones by capturing deeper relationships between code clones beyond syntax \cite{CCLearner, ASTNN, CodeBERTfeng2020codebertpretrainedmodelprogramming,wu2020scdetector,yu2019neural}.
For example, 
Li et al. introduced CCLearner, a clone classifier trained on tokenized data using a deep neural network model\cite{CCLearner}. Zhang et al. proposed ASTNN, which segments abstract syntax trees into sequences of subtrees and encodes them into vectors for code clone detection\cite{ASTNN}. 
Feng et al. proposed CodeBERT, a BERT-based pre-trained model designed to understand and generate both programming and natural languages, with performance improvements achievable through the pretraining-finetuning paradigm\cite{CodeBERTfeng2020codebertpretrainedmodelprogramming}. 
However, several studies have shown that these models exhibit limited generalizability \cite{choi2023investigating,CanNeuralCloneDetectionGeneralizeToUnseenFunctionalitiesƒ}. For instance, Choi et al. found that when CCLearner, ASTNN, and CodeBERT were fine-tuned on one code clone dataset and tested on different datasets, their F1  and MCC-scores dropped significantly \cite{choi2023investigating}. This limitation hinders the practical applicability of these models.

LLM-based code clone detectors have emerged as promising solutions for detecting semantic clones. 
Dou et al. proposed several prompts and evaluated various LLMs' clone detection performance using \textit{BCB}\cite{LLMCCD_ARXIV2023}. Their results showed that \textit{GPT-3.5-turbo} and \textit{GPT-4} outperformed traditional techniques in detecting semantic clones. 
Khajezade et al. compared \textit{GPT-3.5-turbo} with three pre-trained model-based methods (i.e., CodeBERT, RoBERTa, GraphCodeBERT) and found that \textit{GPT-3.5-turbo} achieved the highest recall in \texttt{Java}-\texttt{Java} detection and the highest F1 score in \texttt{Java}-\texttt{Ruby} detection\cite{LLMCCD_ICPC2024}. 
Despite these advancements, research has yet to explore LLM-based clone detectors' generalization and response consistency, which are crucial for their practical application.

\vspace{-1em}
\section{Research Questions}
\label{sec:RQs}
As explained in Section \ref{sec:relatedStudy}, although LLM-based code clone detectors have shown promising results, their generalization across different datasets and response consistency remain underexplored. To address these issues, this study aims to answer the following two RQs:

\textbf{RQ1: How accurately does each LLM detect code clones across different datasets?}
\\ 
This RQ aims to examine LLMs' accuracy in consistently detecting code clones across different datasets. Unlike traditional DL-based code clone detectors that struggle with generalization, LLMs may offer better generalization because they are trained on a much broader and more diverse corpus, which includes code from various sources, languages, and contexts. This potential advantage could significantly enhance their real-world applicability.

\textbf{RQ2: How consistent are the responses from each LLM when presented with identical input?}
 \\ 
This RQ aims to investigate the consistency of results produced by LLM-based code clone detectors when given the same input multiple times.
Due to their non-deterministic nature, LLMs can generate different outputs for the same input. Ensuring high response consistency is crucial for practical applications of LLM-based clone detection. Moreover, reproducibility is essential in software engineering research, as many tasks depend on consistent and reliable clone detection. 
\section{Methodology}
\label{sec:experiments}

This section presents the evaluation methodology used to answer the two RQs discussed in Section \ref{sec:relatedStudy}. Figure \ref{fig:ExperimentProcess} illustrates the overall experimental process. 
First, datasets containing both clone and non-clone code pairs were constructed across four programming languages. Next, each code pair was formatted into a predefined prompt and submitted to the LLMs for code clone detection. Finally, the performance of the LLMs was evaluated using various performance metrics.

\subsection{Clone Datasets}
\textbf{Motivation for Constructing Datasets:}
Existing code clone datasets have two significant limitations: they lack diversity in programming languages beyond \texttt{Java} and do not account for syntactic similarity when selecting clone pairs. 
We constructed seven diverse code clone datasets to address these limitations and answer RQ1 and RQ2. These datasets cover a range of programming languages and syntactic similarities, ensuring they better reflect the diverse types of code clones developers encounter in practice. The primary motivations behind constructing these datasets were to incorporate syntactic similarity as a key factor influencing the complexity of code clone detection and ensure fair comparisons across LLMs by controlling the distribution of syntactic similarity within the datasets. By carefully curating clone pairs with varying degrees of similarity, we could report F1 scores for the LLMs across different similarity levels, providing valuable insights into their performance.

\textbf{Dataset Construction Methodology:}
Rather than random sampling, we selected clone pairs based on their similarity using the Levenshtein ratio\cite{LDistanceRatio} on identifier-normalized token sequences, which is defined by the following equation:
\begin{equation}
    \label{equ:LevSimi}
    LevRatio(A,B) = \frac{Max\left ( \left | A \right |,  \left | B \right |\right ) - LevTS\left (A,B\right )}{Max\left ( \left | A \right |,  \left | B \right |\right )} 
\end{equation}

Here, $LevTS(A, B)$ represents the Levenshtein distance between two token sequences, $A$ and $B$, with identifiers normalized \cite{LDistance}.
We chose this similarity measurement for two key reasons. 
It enables the creation of balanced datasets by ensuring an even distribution of clone pairs across different similarity levels.
The Levenshtein ratio ensures that semantic clones are spread across the similarity range, whereas syntactic clones tend to cluster near 100\% similarity. 
Second, it was also adopted by Zhu et al. \cite{MSCCD_JSS} to categorize code pairs within the \textit{CN}. By using the same established approach, we enhanced the reliability and consistency of our similarity measurements across a wide range of levels, thus strengthening the robustness of our evaluation.

\textbf{Constructed Datasets}: 
We constructed seven datasets, in which \textbf{\textit{CN-random}} is the primary dataset for most experiments. The other six datasets allow for comparisons of LLM performance across different datasets, similarity ranges, and programming languages.

\begin{itemize}
    \item \textbf{\textit{CN-random}} consists of 700 clone pairs and 700 non-clone pairs in \texttt{Java}, which were randomly sampled from 14 subsets of Zhu et al.'s dataset  \cite{MSCCD_JSS}.
     Clone pairs originate from code submissions to the same programming competition problems, while non-clone pairs were sampled from different problems. Clones in this dataset were distributed across 20\% similarity intervals, with 10 pairs randomly selected from each range based on the similarity distribution. Non-clone pairs were evenly sampled from different subsets to ensure diversity.

    \item \textbf{\textit{BCB-random}} was constructed from \textit{BCB} to have identical similarity distrubutions with \textit{CN-random}, ensuring comparable distributions across datasets. 
    Since \textit{BCB} uses a different similarity measure than that of \textit{CN}, clone pairs in \textit{BCB} generally have lower similarity values. To address this problem, we adjusted the similarity ranges to match those used in \textit{CN}, recalculated the similarity for each code pair, and resampled any pairs that did not satisfy the required similarity threshold. 

    \item \textbf{\textit{CN-average}} was constructed to compute F1 scores within each similarity range, including an equal number of clone and non-clone pairs in each range. Initially, we focus on constructing a single dataset for all experiments, avoiding the necessity of separating sampling strategies. However, obtaining sufficient non-clones for similarity ranges above 60\% was challenging. Therefore, for similarity ranges below 0.6, we sampled 140 clone and non-clone pairs. For ranges above 0.6, we sampled only 140 clone pairs. Recall metrics are reported exclusively for the ranges that contain only clone pairs.

    \item \textbf{\textit{BCB-average}}, similar to \textit{CN-average}, includes an equal number of clones and non-clones pairs in higher similarity ranges (above 0.6). This reflects the higher proportion of high-similarity non-clones in \texttt{BCB}.

    \item \textbf{\textit{CN-random-C}}, \textbf{\textit{CN-random-C++}}, \textbf{\textit{CN-random-Python}} were constructed using the same sampling strategy with \textit{CN-random}, but include clone pairs in \texttt{C}, \texttt{C++}, and \texttt{Python}, respectively. Similar to \texttt{CN-random}, clones in these datasets were distributed across 20\% similarity intervals, with 10 pairs randomly selected from each range. Non-clone pairs were sampled evenly from different subsets to maintain diversity.
\end{itemize}

Table \ref{Tab:DetailsOfJavaDatasets} presents the number of clone and non-clone pairs collected within each Levenshtein ratio range in all the constructed datasets.
As depicted in this table, \textit{CN-random} and \textit{BCB-random} have identical distributions, whereas \textit{CN-average} and \textit{BCB-average} contain an equal number of clones and non-clones pairs across as many similarity ranges as possible.

\subsection{Models and Prompting}
\label{Sec:ModelsAndPrompting}
This study evaluated five LLMs, including a reasoning model named \textit{o3-mini}, two models from the GPT-series (\textit{GPT-4o}, and \textit{GPT-4o-mini}), and two open-source LLMs, \textit{Llama 3.1} and \textit{Mistral}.

\begin{itemize}

\item  \textbf{\textit{o3-mini}} is one of the latest reasoning models of OpenAI. Reasoning models produce a long internal chain of thought before their response and aim at more complex tasks such as coding and scientific reasoning. \textit{o3-mini} was selected to investigate the performance of the reasoning model in clone detection.

\item \textbf{\textit{GPT-4o}} is one of the latest and most advanced models in the GPT series. This study uses \textit{GPT-4o-2024-08-06}, which is the most recent available at the time. We selected this model to investigate whether the advancements in the model can enhance performance in code clone detection.
\item \textbf{\textit{GPT-4o-mini}} is the lightweight version of \textit{GPT-4o}. Evaluating this model allows us to assess whether the mini version exhibits performance degradation in the code clone detection task. \textit{GPT-4o-mini} points to \textit{GPT-4o-mini-2024-07-18}. We used this model to examine the trade-off between model size and performance, particularly in code clone detection.
\item \textbf{\textit{Llama 3.1:8B}}\footnote{Model ID in oLlama platform: 42182419e950} is an open-source LLM released by Meta. We chose this model because it provides a lower-cost alternative to proprietary models, and understanding its performance can provide insights into whether open-source models can match or outperform commercial alternatives. The specific version in this study is \textit{Llama 3.1:8B}, which has 8 billion parameters.

\item \textbf{\textit{Mistral:7B}}\footnote{Model ID in oLlama platform: f974a74358d6} is another open-source LLM released by Mistral AI, but with a smaller parameter size (7 billion parameters). \textit{Mistral} was chosen to evaluate whether smaller open-source models can still perform well compared with larger commercial models. The version used in this study is \textit{Mistral:7B}.
\end{itemize}

We utilized Batch API\footnote{\url{https://platform.openai.com/docs/guides/batch}} of Open AI to access their models, while \textit{Llama 3.1} and \textit{Mistral} were run locally using  oLlama\footnote{\url{https://oLlama.com/}}.
All models were evaluated in their original, non-fine-tuned form.

We employed four distinct prompts to conduct code clone detection with these LLMs, as detailed in Table \ref{Tab:PromptsList}. The first prompt, \textit{P0}, was identified as the optimal prompt in the study by Khajezade et al. \cite{LLMCCD_ICPC2024}. 
The remaining prompts, \textit{P1}, \textit{P2}, and \textit{P3}, are sourced from the study by Dou et al.,. In their original study, \textit{P3} is a prompt with relatively high recall, \textit{P2} is a prompt with relatively high precision, and \textit{P1} is a prompt that balances recall and precision \cite{LLMCCD_ARXIV2023}. Prompt \textit{P0} asks LLMs to judge if two codes solve identical problems, while the other three prompts ask LLMs to assess the code pair's similarity to make the judgment.

\subsection{Performance Evaluation for Code Clone Detection}

In this study, the performance of LLMs for code clone detection was evaluated based on their ability to provide binary responses, specifically `yes' or `no'. To automate the collection of results of code clone detection, each selected prompt, described in Section \ref{Sec:ModelsAndPrompting}, was designed to instruct the LLMs to return only these two responses. However, LLMs occasionally deviated from this instruction by including additional text, such as reasoning, in their responses. To address this problem, a discriminator program was implemented to process and convert the responses into Boolean results. The discriminator accepts variations such as ``Result: Yes'' or ``Yes, …'' but excludes more complex responses, even if they contain the correct answer, to avoid recognition problems.

The temperature parameter (\textit{temp}) is a critical factor in determining the randomness of LLMs' output. Lower \textit{temp} values generally lead to more deterministic and consistent responses, whereas higher \textit{temp} values introduce greater variability in the output. In RQ1, we set the  \textit{temp} value as 0.3 to ensure consistency with the existing study \cite{LLMCCD_ICPC2024}. As a relatively low value, setting \textit{temp} to 0.3 is considered to balance answer consistency and detection ability. 
In RQ2, we measured LLMs' response consistency rate and F1 score for various \textit{temp} settings. Because the effect of \textit{temp} depends on the model's architecture, \textit{temp} values are incomparable across different LLMs. Besides, \textit{o3-mini} does not support the temperature parameter.

For performance evaluation, the study uses the commonly adopted metrics of recall, precision, and F1 score. Recall is calculated as $\frac{TP}{TP + FN}$, whereas precision is defined as $\frac{TP}{TP + FP}$. F1 score (harmonic mean of recall and precision) is computed as $\frac{2 \times Recall \times Precision}{Recall + Precision}$. Here, $TP$, $FP$, $TN$, and $FN$ represent the true positives, false positives, true negatives, and false negatives, respectively. 
Specifically, $TP$ refers to the intersection of true clones and detected clones, $FP$ is the intersection of false clones and detected clones, $TN$ refers to the intersection of false clones and non-detected clones, and $FN$ is the intersection of true clones and non-detected clones.

To address RQ2, we evaluate the response consistency rate (RCR) of LLMs for code clone detection. This metric reflects the models’ ability to provide consistent outputs for identical inputs across repeated submissions. The response consistency rate for $i$-th submission $RCR_{i}$ is defined in Equation \ref{Equ:RCR}. 
\begin{equation}
{\fontsize{6}{0}\selectfont
\label{Equ:RCR}
\text{RCR}_i = \frac{1}{N}\sum_{j=1}^{N}\prod_{k=2}^{i}\mathbf{1}\{r_k(j)=r_1(j)\}, 
\mathbf{1}\{A\} =
\begin{cases}
1, & A \text{ is true}\\[1mm]
0, & \text{otherwise}
\end{cases}
}
\end{equation}
Here, $r_{k}(j)$ represents the judgment for the $j$-th code pair in the $k$-th submission. Given that there are $N$ code pairs in the test dataset, $RCR_{i}$ is defined as the proportion of cases in that submission where the judgment has remained unchanged with the first submission.

\begin{figure*}[t]
\centering
    \graphicspath{{./img/}}

    \begin{subfigure}{\textwidth}
        \includegraphics[width=\textwidth]{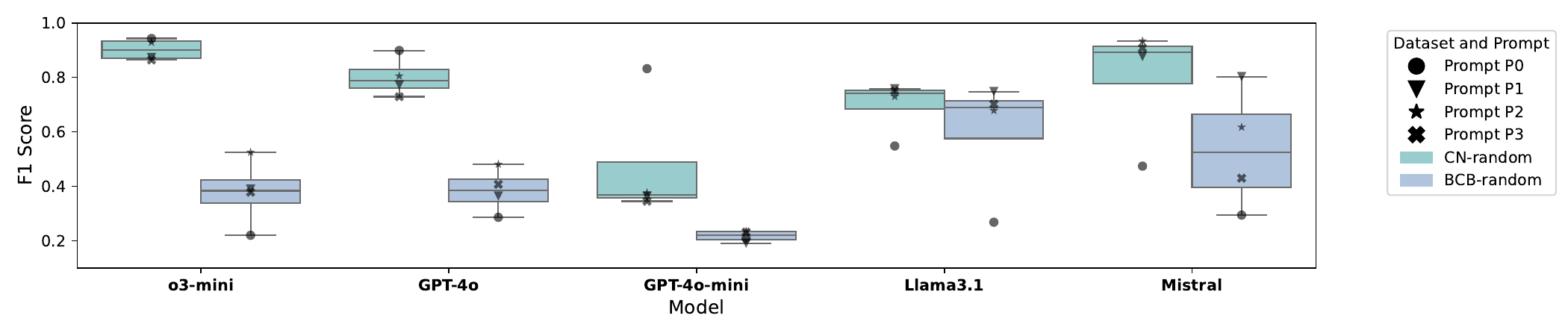}
        \caption{Results across \textit{CN-random} and \textit{BCB-random}}
        \label{fig:RQ2-d}
    \end{subfigure}

    \begin{subfigure}{\textwidth}
        \includegraphics[width=\textwidth]{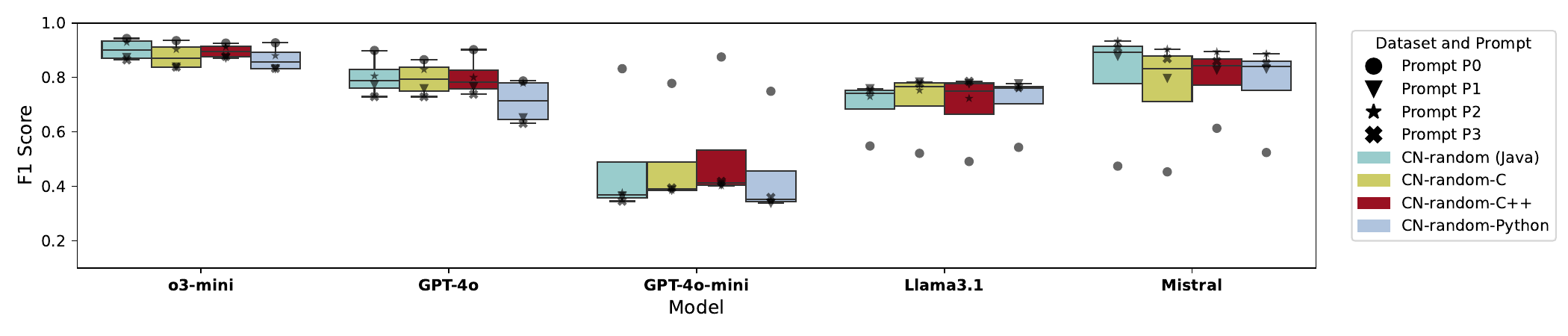}
        \caption{Results across Four Languages of \textit{CN-random}}
        \label{fig:RQ2-l}
    \end{subfigure}

    \caption{F1-Score of Each Model and Prompt for Different Datasets}
    \label{fig:RQ2-2}
\end{figure*}

\section{Answers to the RQs}
\label{sec:results}

\subsection{RQ1: How accurately does each LLM detect code clones across different datasets?}
\label{sec:RQ1}

To address this RQ, we conducted experiments on our constructed datasets to evaluate the code clone detection performance of various LLMs. First, we analyzed the performance on \textit{CN-random}, the main dataset used in almost all experiments, to assess LLMs' ability to detect clones in a dataset with diverse similarity ranges. To further investigate the impact of dataset characteristics, we compared the detection performance of LLMs on \textit{CN-random} and \textit{BCB-random}, which share identical similarity distributions but originate from different code collections.  
This comparison helped us examine whether models maintain consistent performance across datasets.
Furthermore, we assessed the performance on datasets covering four programming languages from \textit{CN} to determine whether models generalize well across different languages. Finally, we compared LLM's performance on \textit{CN-average} and \textit{BCB-average} to analyze detection accuracy across each similarity range.

\begin{table}[t]
\centering
\caption{Results of Each Prompt and Model for \textit{CN-random}}
\renewcommand{\arraystretch}{0.4}

\label{Tab:ResRQ2_1}
\resizebox{0.48\textwidth}{!}{

\begin{tabular}{m{1.3cm}cccc}
\toprule
Model & PromptId & Recall & Precision & F1 \\ \midrule
\multirow{4}{*}[-1.2em]{\textit{o3-mini}} & \textit{P0} & 0.891 & 1.000 & \textbf{0.943} \\ \cmidrule(lr){2-5} 
   & \textit{P1} & 0.774 & 1.000 & 0.873 \\ \cmidrule(lr){2-5} 
   & \textit{P2} & 0.873 & 0.993 & 0.929 \\ \cmidrule(lr){2-5} 
   & \textit{P3} & 0.760 & 1.000 & 0.864 \\ \midrule
\multirow{4}{*}[-1.2em]{\textit{GPT-4o}} & \textit{P0} & 0.817 & 1.000 & 0.899 \\ \cmidrule(lr){2-5} 
   & \textit{P1} & 0.630 & 0.998 & 0.772 \\ \cmidrule(lr){2-5} 
 & \textit{P2} & 0.677 & 0.994 & 0.805 \\ \cmidrule(lr){2-5} 
   & \textit{P3} & 0.574 & 0.998 & 0.729 \\ \midrule 
\multirow{4}{*}[-1.2em]{\textit{GPT-4o-mini}} & \textit{P0} & 0.714 & 0.996 & 0.832 \\ \cmidrule(lr){2-5} 
   & \textit{P1} & 0.221 & 1.000 & 0.363 \\ \cmidrule(lr){2-5} 
   & \textit{P2} & 0.231 & 1.000 & 0.376 \\ \cmidrule(lr){2-5} 
   & \textit{P3} & 0.209 & 1.000 & 0.345 \\ \midrule 
\multirow{4}{*}[-1.2em]{\textit{Mistral}} & \textit{P0} & 0.311 & 0.995 & 0.474 \\ \cmidrule(lr){2-5} 
   & \textit{P1} & 0.990 & 0.787 & 0.877 \\ \cmidrule(lr){2-5} 
   & \textit{P2} & 0.938 & 0.929 & \textbf{0.934} \\ \cmidrule(lr){2-5} 
   & \textit{P3} & 0.862 & 0.958 & 0.908 \\ \midrule 
\multirow{4}{*}[-1.2em]{\textit{Llama3.1}} & \textit{P0} & 0.384 & 0.954 & 0.548 \\ \cmidrule(lr){2-5} 
   & \textit{P1} & 0.997 & 0.611 & 0.758 \\ \cmidrule(lr){2-5} 
   & \textit{P2} & 0.889 & 0.619 & 0.730 \\ \cmidrule(lr){2-5} 
   & \textit{P3} & 0.794 & 0.713 & 0.751 \\ 
\bottomrule
\end{tabular}

}
\end{table}

\noindent
\textbf{Results on the Main Dataset: \textit{CN-random}}

The results of code clone detection on  \textit{CN-random} are listed in Table \ref{Tab:ResRQ2_1}. 
Among all models, \textit{o3-mini} achieved the highest overall F1 score of 0.943 when using prompt \textit{P0} from Table \ref{Tab:PromptsList}.
\textit{Mistral} also achieved a 0.934 F1 score in prompt \textit{P2}.
The \textit{GPT-4o} and \textit{GPT-4o-mini} also performed well, with their best F1 scores of 0.899 and 0.832, respectively. 
In contrast, \textit{Llama 3.1} exhibited the lowest performance, with a maximum F1 score of 0.758. 
Additionally, we observed the following two findings:
\begin{itemize}
 
    \item \textbf{Performance variation across prompts for the same model}: For all five models, the difference between the highest and lowest F1 scores across the four prompts was substantial, with the slightest difference being 0.17. \textit{GPT-4o-mini} and \textit{Mistral} revealed even more significant differences, with gaps of 0.49 and 0.46, respectively.
    \item \textbf{Performance variation across models for the same prompt}: Prompt \textit{P0}, which achieved F1 scores above 0.8 with the GPT-series models, yielded only 0.548 and 0.474 F1 scores with \textit{Llama 3.1} and \textit{Mistral}. In contrast, prompt \textit{P2}, which produced the highest overall F1 score with \textit{Mistral} (0.934), achieved only 0.376 with \textit{GPT-4o-mini}.
\end{itemize}
These results highlight that the choice of prompt significantly impacts model performance, suggesting that selecting the optimal model-prompt combination is more crucial than simply choosing the most powerful model.

\noindent
\textbf{Comparison Between \textit{CN-random} and \textit{BCB-random}}

To evaluate the impact of dataset differences on clone detection, we compared model results between \textit{CN-random} and \textit{BCB-random}. 
Figure \ref{fig:RQ2-2}(\subref{fig:RQ2-d}) presents a comparison of the F1 score for \textit{CN-random} and \textit{BCB-random}, revealing that all models achieved considerably lower F1 scores on \textit{BCB-random} than \textit{CN-random}. For each model, the F1 score across all prompts on \textit{BCB-random} was consistently lower than on \textit{CN-random}. 
To quantify these differences, Table \ref{tab:MeanDiffF1Score} presents the F1 score differences between the two datasets for each prompt. The most significant relative drop was observed with \textit{o3-mini}, where the average F1 score decreased by 0.52. In contrast, only four model-prompt combinations had F1 score differences smaller than 0.1. The smallest relative drop was observed with \textit{Llama 3.1}, with an average F1 score decrease of only 0.1 across all prompts. These results demonstrated that most model-prompt combinations could not achieve the same level of performance on \textit{CN-random} and \textit{BCB-random}, indicating the challenge of achieving consistent performance across different code collections, even when similarity distributions are aligned.

\begin{table}[]
\centering
\caption{Mean difference of F1 score for \textit{CN-random} and \textit{BCB-random} \\ (Difference of F1 score: F1 score for \textit{CN-random} - F1 score for \textit{BCB-random})}
\label{tab:MeanDiffF1Score}
\renewcommand{\arraystretch}{0.4}

{
\resizebox{0.49\textwidth}{!}{
\begin{tabular}{lccccc}
\toprule
\multirow{2}{*}{Model} & \multicolumn{4}{c}{Difference of F1 score}                                                                       & \multirow{2}{*}{Mean} \\ \cmidrule(lr){2-5}
                       & \multicolumn{1}{c}{\textit{P0}} & \multicolumn{1}{c}{\textit{P1}} & \multicolumn{1}{c}{\textit{P2}} & \textit{P3} &                       \\ \midrule
\textit{o3-mini}          & \multicolumn{1}{c}{0.72}       & \multicolumn{1}{c}{0.48}       & \multicolumn{1}{c}{0.40}       & 0.49       & \textbf{0.52}                  \\ \midrule
\textit{GPT-4o}                 & \multicolumn{1}{c}{0.61}       & \multicolumn{1}{c}{0.41}       & \multicolumn{1}{c}{0.32}       & 0.32       & 0.42                  \\ \midrule
\textit{GPT-4o-mini}            & \multicolumn{1}{c}{0.62}       & \multicolumn{1}{c}{0.17}       & \multicolumn{1}{c}{0.14}       & 0.11       & 0.26                  \\ \midrule
\textit{Llama 3.1}               & \multicolumn{1}{c}{0.28}       & \multicolumn{1}{c}{0.01}       & \multicolumn{1}{c}{0.05}       & 0.05       & 0.10                  \\ \midrule
\textit{Mistral}                & \multicolumn{1}{c}{0.18}       & \multicolumn{1}{c}{0.07}       & \multicolumn{1}{c}{0.32}       & 0.48       & 0.26                  \\ \midrule \midrule
Mean                   & \multicolumn{1}{c}{0.44}       & \multicolumn{1}{c}{0.17}       & \multicolumn{1}{c}{0.22}       & 0.26       &                       \\ \bottomrule
\end{tabular}
}
}
\end{table}

\noindent
\textbf{Comparison Among \textit{CN-random} in Four Languages}

Figure \ref{fig:RQ2-2}(\subref{fig:RQ2-l}) presents the results of each LLM for four languages of \textit{CN-random}. The four boxes from left to right for each model correspond to \texttt{Java}, \texttt{C}, \texttt{C++}, and \texttt{Python}.  Unlike the cross-dataset comparison, all models exhibited similar performance across the four programming languages. This indicates that the models can generalize well across different programming languages within the \textit{CN-random} datasets, maintaining stable performance. Thus, language-specific variations within \textit{CN} have a more negligible impact on performance than the differences between \textit{CN} and \textit{BCB}.

\begin{figure*}[]
    \centering
    \graphicspath{{./img/}}
        \includegraphics[width=\textwidth,height=430px]{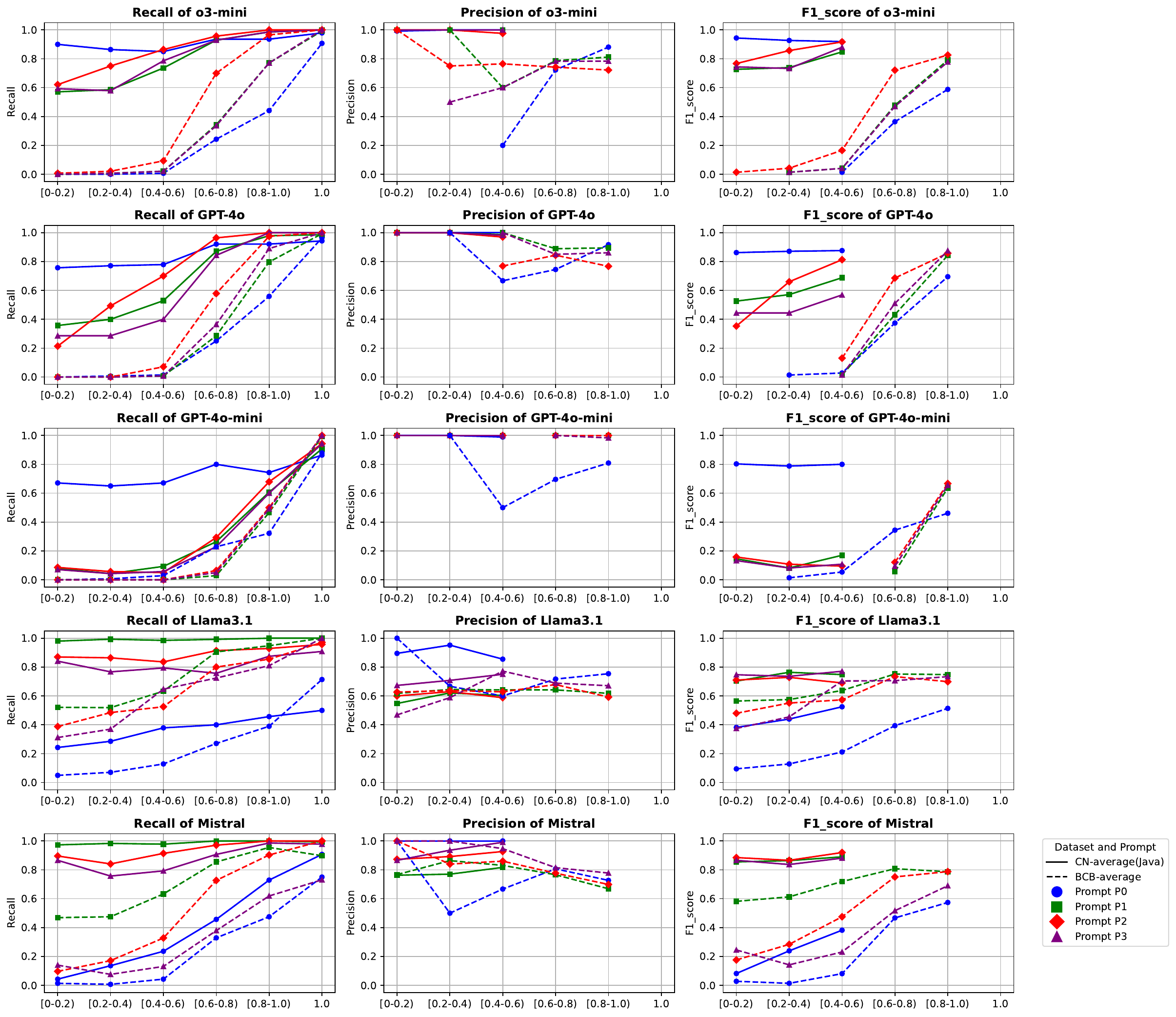}
    \caption{Recall, Precision, and F1-Score of Each Model and Prompt in Each Similarity Range}
    \label{fig:RQ3}
\end{figure*}

\noindent
\textbf{Comparison Between \textit{CN-average} and \textit{BCB-average}}

Figure \ref{fig:RQ3} presents recall, precision, and F1 scores across different similarity ranges, solid lines representing \textit{CN-average} results, and dashed lines representing \textit{BCB-average} results. 
In this figure, different prompts are distinguished by color and marker type.
For \textit{CN-average}, F1 scores are reported for similarity ranges lower than 0.6, whereas for \textit{BCB-average}, the ranges are reported for ranges lower than 1.0.
In the other ranges, only recall is reported due to the limited number of non-clones available. Furthermore, when all candidate code pairs were classified as non-clones, precision and F1 score could not be calculated (e.g., \textit{o3-mini} with prompt \textit{P0} in the second range of \textit{BCB-average}).

The results indicate that higher similarity code clones are easier to detect, aligning with expectations. Most model-prompt combinations exhibit higher recall for code pairs with higher similarity. Furthermore, all models demonstrate significantly higher recall for \textit{CN-average} compared to \textit{BCB-average}. In areas where code similarity is below 0.8, nearly all model-prompt combinations show a noticeably higher recall for \textit{CN-average} than for \textit{BCB-average}.  However, an exception is observed in \textit{GPT-4o-mini}, where other prompts yield similar recall across both datasets in all regions except for Prompt \textit{P0}. It is also noteworthy that some prompt-model combinations fail to achieve 1.0 recall even in regions where similarity is 1.0 (T1 and T2 clones). For instance, Prompt \textit{P0} does not attain 1.0 recall in this region across any model. This may be due to Prompt \textit{P0} emphasizing functional equivalence in code clones, leading to Type-2 clones with different function names to be misclassified as non-clones. Additionally, recall for Prompt \textit{P0} varies significantly across models, with recall on \textit{GPT-4o} nearly twice that on \textit{Llama} 3.1. This result suggests that selecting the most suitable prompt is essential for achieving optimal performance across different models.

Another notable observation is that non-clones with high similarity are more likely to be misclassified as clones. However, the precision of various models across high- and low-similarity code pairs does not show as large a disparity as recall does. Each model demonstrates cases where precision is higher in regions of higher similarity. Overall, the models maintained relatively high precision, with all model-prompt combinations achieving at least 0.6 precision in the most challenging area. Additionally, in comparable areas, most model-prompt combinations achieve precision for \textit{CN-average} that is equal to or greater than for \textit{BCB-average}. F1 scores correlate positively with code similarity, primarily due to their dependence on recall. Within comparable similarity regions, all model-prompt combinations exhibit significantly higher F1 scores for \textit{CN-average} than for \textit{BCB-average}.

\begin{table*}[t]
\captionsetup{justification=centering}
\caption{Max Recall and Precision in Each Similarity Range. \\ For similarity ranges lower than 0.8, almost all LLMs' max recall for \textit{CN-average} is considerably higher than that for \textit{BCB-average}. }
\label{tab:maxRecallInSimiRange}
\renewcommand{\arraystretch}{0.4}

\resizebox{0.98\textwidth}{!}{
{

\begin{tabular}{lc|cccccc|ccccc}
\toprule
\multirow{2}{*}{Model}         & \multirow{2}{*}{Dataset} & \multicolumn{6}{c}{Recall}                                                                                                                                                            & \multicolumn{5}{|c}{Precision}                                                                                                                        \\ \cmidrule(lr){3-13} 
                               &                          & \multicolumn{1}{c}{{[}0,0.2)} & \multicolumn{1}{c}{{[}0.2,0.4)} & \multicolumn{1}{c}{{[}0.4,0.6)} & \multicolumn{1}{c}{{[}0.6,0.8)} & \multicolumn{1}{c}{{[}0.8,1{)}} & {[}1,1{]} & \multicolumn{1}{c}{{[}0,0.2)} & \multicolumn{1}{c}{{[}0.2,0.4)} & \multicolumn{1}{c}{{[}0.4,0.6)} & \multicolumn{1}{c}{{[}0.6,0.8)} & {[}0.8,1{)} \\  \midrule 
\multirow{2}{*}{\textit{o3-mini}} & \textit{CN-average}               & \multicolumn{1}{c}{0.900}     & \multicolumn{1}{c}{0.864}       & \multicolumn{1}{c}{0.864}       & \multicolumn{1}{c}{0.957}       & \multicolumn{1}{c}{1.000}       & 1.000      & \multicolumn{1}{c}{1.000}       & \multicolumn{1}{c}{1.000}         & \multicolumn{1}{c}{1.000}       & \multicolumn{1}{c}{-}           & -           \\ \cmidrule(lr){2-13} 
                               & \textit{BCB-average}              & \multicolumn{1}{c}{0.007}     & \multicolumn{1}{c}{0.021}       & \multicolumn{1}{c}{0.093}       & \multicolumn{1}{c}{0.700}       & \multicolumn{1}{c}{0.966}       & 1.000      & \multicolumn{1}{c}{1.000}       & \multicolumn{1}{c}{1.000}       & \multicolumn{1}{c}{0.765}       & \multicolumn{1}{c}{0.787}        & 0.881        \\ \midrule \midrule
\multirow{2}{*}{\textit{GPT-4o}}        & \textit{CN-average}               & \multicolumn{1}{c}{0.757}     & \multicolumn{1}{c}{0.771}       & \multicolumn{1}{c}{0.779}       & \multicolumn{1}{c}{0.921}       & \multicolumn{1}{c}{0.964}       & 1.000      & \multicolumn{1}{c}{1.000}       & \multicolumn{1}{c}{1.000}         & \multicolumn{1}{c}{1.000}         & \multicolumn{1}{c}{-}           & -           \\ \cmidrule(lr){2-13} 
                               & \textit{BCB-average}              & \multicolumn{1}{c}{0}         & \multicolumn{1}{c}{0.007}       & \multicolumn{1}{c}{0.071}       & \multicolumn{1}{c}{0.579}       & \multicolumn{1}{c}{0.975}       & 1.000      & \multicolumn{1}{c}{NaN}       & \multicolumn{1}{c}{1.000}         & \multicolumn{1}{c}{1.000}         & \multicolumn{1}{c}{0.890}        & 0.917       \\ \midrule \midrule
\multirow{2}{*}{\textit{GPT-4o-mini}}   & \textit{CN-average}               & \multicolumn{1}{c}{0.671}     & \multicolumn{1}{c}{0.650}        & \multicolumn{1}{c}{0.671}       & \multicolumn{1}{c}{0.800}         & \multicolumn{1}{c}{0.743}       & 0.943     & \multicolumn{1}{c}{1.000}       & \multicolumn{1}{c}{1.000}         & \multicolumn{1}{c}{1.000}         & \multicolumn{1}{c}{-}           & -           \\ \cmidrule(lr){2-13} 
                               & \textit{BCB-average}              & \multicolumn{1}{c}{0}         & \multicolumn{1}{c}{0.007}       & \multicolumn{1}{c}{0.029}       & \multicolumn{1}{c}{0.229}       & \multicolumn{1}{c}{0.500}         & 1.000      & \multicolumn{1}{c}{NaN}       & \multicolumn{1}{c}{1.000}         & \multicolumn{1}{c}{0.500}         & \multicolumn{1}{c}{1.000}         & 1.000        \\ \midrule \midrule
\multirow{2}{*}{\textit{Llama 3.1}}      & \textit{CN-average}               & \multicolumn{1}{c}{0.980}      & \multicolumn{1}{c}{0.993}       & \multicolumn{1}{c}{0.985}       & \multicolumn{1}{c}{0.992}       & \multicolumn{1}{c}{1.000}         & 1.000      & \multicolumn{1}{c}{0.895}     & \multicolumn{1}{c}{0.952}       & \multicolumn{1}{c}{0.855}       & \multicolumn{1}{c}{-}           & -           \\ \cmidrule(lr){2-13} 
                               & \textit{BCB-average}              & \multicolumn{1}{c}{0.521}     & \multicolumn{1}{c}{0.519}       & \multicolumn{1}{c}{0.647}       & \multicolumn{1}{c}{0.905}       & \multicolumn{1}{c}{0.948}       & 1.000      & \multicolumn{1}{c}{1.000}       & \multicolumn{1}{c}{0.667}       & \multicolumn{1}{c}{0.722}       & \multicolumn{1}{c}{0.717}       & 0.754       \\ \midrule \midrule
\multirow{2}{*}{\textit{Mistral}}       & \textit{CN-average}               & \multicolumn{1}{c}{0.973}     & \multicolumn{1}{c}{0.983}       & \multicolumn{1}{c}{0.978}       & \multicolumn{1}{c}{1.000}         & \multicolumn{1}{c}{1.000}         & 1.000      & \multicolumn{1}{c}{1.000}       & \multicolumn{1}{c}{1.000}         & \multicolumn{1}{c}{1.000}         & \multicolumn{1}{c}{-}           & -           \\ \cmidrule(lr){2-13} 
                               & \textit{BCB-average}              & \multicolumn{1}{c}{0.469}     & \multicolumn{1}{c}{0.475}       & \multicolumn{1}{c}{0.633}       & \multicolumn{1}{c}{0.855}       & \multicolumn{1}{c}{0.955}       & 1.000      & \multicolumn{1}{c}{1.000}       & \multicolumn{1}{c}{1.000}         & \multicolumn{1}{c}{0.947}       & \multicolumn{1}{c}{0.815}       & 0.778       \\ 
\bottomrule
\end{tabular}
}
}

\end{table*}

To further examine model performance, Table \ref{tab:maxRecallInSimiRange} summarizes the highest recall and precision achieved by each model. The highest recall and precision in each range for a given model may not be achieved by the same prompt. For \textit{CN-average}, \textit{Llama 3.1} and \textit{Mistral} achieve recall above 0.97 even in similarity ranges lower than 0.6, considerably outperforming the GPT-series in these ranges. \textit{Mistral} demonstrated slightly higher precision than \textit{Llama 3.1}, making it the best-performing model on \textit{CN-average}. Similarly, on \textit{BCB-average}, although \textit{Llama 3.1} achieved a slightly higher recall, \textit{Mistral} demonstrated a more pronounced advantage in precision. As a result, \textit{Mistral} emerged as the best-performing model for \textit{BCB-average}, demonstrating a strong balance between recall and precision.

    \noindent 
    \fbox{
        \parbox{0.465\textwidth}{
      The answer to RQ1:
None of the models achieved comparable detection capability across datasets derived from different code collections. 
        }
      }

\subsection{RQ2: How consistent are the responses from each LLM when presented with identical input?}
\label{sec:RQ2}

We evaluated the response consistency of each LLM by submitting the same input multiple times using the \textit{CN-random} dataset.
In common with most DL models, LLMs exhibit non-determinism, often yielding variable outputs for identical inputs.
This behavior is particularly sensitive to the input's position relative to the model's decision boundaries for that task.
When the input is situated far away from decision boundaries, the resulting output is typically stable and consistent. Conversely, inputs that lie near a boundary are more likely to produce divergent or inconsistent results.
In clone detection, such inconsistencies reduce reproducibility, undermining the reliability of detection results and limiting the model's practical applicability.
For this experiment, we submitted the same prompt five times per model and recorded the results. Each submission was conducted independently to avoid mutual influence. Additionally, we tested two \textit{temperature} (\textit{temp}) settings: 0.3 (the lowest value) and the maximum value of each model.

\begin{figure*}[]
    \centering
    \graphicspath{{./img/}}

        \includegraphics[width=\textwidth,height=225px]{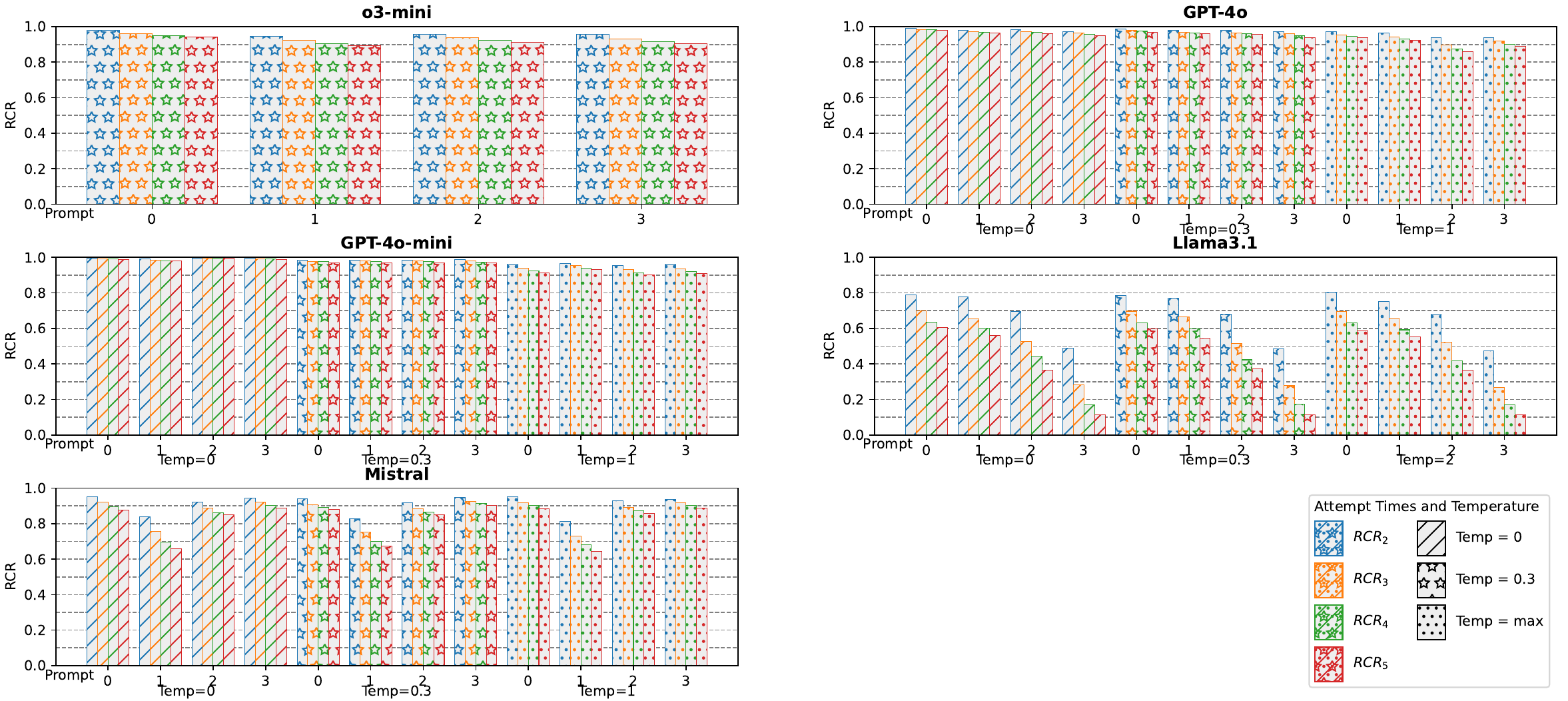}
    \caption{Response Consistency Rate (RCR) of Each Model}
    \label{fig:RQ4_C}
\end{figure*}

Figure \ref{fig:RQ4_C} shows the proportion of responses that remained consistent with the first submission across the second, third, fourth, and fifth trials for each model-prompt combination. As shown in the figure, \textit{GPT-4o} and\textit{GPT-4o-mini} demonstrated the highest consistency, maintaining a rate above 90\% in nearly all cases. 
\textit{Mistral} exhibited considerable variation in consistency depending on the prompt, with prompt \textit{P1} showing notably lower consistency than the others, typically lower than 0.8. 
\textit{Llama 3.1}, on the other hand, exhibited the lowest consistency. By the second submission, at least 20\% of responses differ from the first trial across all prompts. By the fifth submission, at least 40\% of responses were inconsistent, with prompt \textit{P3} showing extreme instability, reaching 80\% inconsistency.

To analyze the factors influencing response consistency, we examine how prompt selection and  \textit{temp}  settings affected consistency variation. Table \ref{tab:RQ4:ConsistencyRateRange} presents the results. 
The average consistency rate range affected by \textit{temp} was calculated by finding the difference in consistency between \textit{temp} 0 and the maximum \textit{temp} for each prompt submission and subsequently averaging the results. 
The average consistency rate range affected by the prompt was calculated by determining the range of consistency rates across four prompts and averaging the results. 
Overall, \textit{temp} had minimal impact on consistency, with fluctuations generally below 0.08 across all models.
However, prompt choice significantly influenced consistency, particularly for \textit{Llama 3.1} and \textit{Mistral}, which showed large variations exceeding 0.4 across different prompts. For most models, while changes in \textit{temp} had little effect on consistency, choosing the right prompt played a critical role in maintaining stability.

Moreover, we calculated the range of F1 scores achieved by each model under various \textit{temp} settings to assess how inconsistent responses affected overall model performance. These results are presented in Table \ref{tab:averageF1Range}. For all models, F1 score fluctuations were lower than 0.032, and the degree of F1 score variation did not significantly differ across \textit{temp} settings. This result indicates that \textit{temp} settings do not significantly affect the performance of LLMs in code clone detection.

One exception in this study was \textit{o3-mini}, which does not allow \textit{temp} configuration. As a result, we submitted each prompt five times under its default settings. \textit{o3-mini} maintained a response consistency rate above 0.9 for all four prompts by the fifth submission. Moreover, the differences between prompts were minimal, with response consistency rates fluctuating by only 0.040. Similarly, the variation in F1 scores across different prompts was slight, with an average F1 fluctuation range of just 0.012. These results indicate that \textit{o3-mini} is a highly stable model in code clone detection, maintaining consistent responses regardless of prompt choice.
    \noindent  
    \fbox{
        \parbox{0.465\textwidth}{
The answer to RQ2: Most models' response consistency is higher-than-expected, and their inconsistency did not result in a significant fluctuation in the detection capability overall. 
    }
 }

\section{Discussions}
\label{sec:Discussions}

\subsection{Toward Real-world Applications of LLM-based Clone Detection}

This section discusses the challenges of using LLM-based clone detection in practical applications. In Section \ref{sec:results}, we evaluated five LLMs' accuracy and response consistency. Although each LLM exhibits high detection accuracy in CN-related datasets, the accuracy of each model in BCB-related datasets declined considerably. These results indicate that the high accuracy of these LLMs for competitive programming data cannot be reflected in open-source software. Competitive programming data has clear labels compared to open-source software, rendering learning easy for large models. In this study, we did not use fine-tuning or prompt-tuning. In future research, we should actively use these technologies to improve the accuracy of clone code detection in open-source software by large models. Regarding answer consistency, our experimental results revealed pronounced differences between models. When choosing an LLM, we should focus on the difference in answer consistency. Furthermore, the cost of using LLMs should be considered. The cost of LLMs can be categorized into time and economic costs. Regarding running time, LLMs typically require very high computing resources; otherwise, the response speed decreases considerably. Communication time costs also occur when using commercial models through APIs. Economically, this also includes the cost of using commercial models. In practice, existing clone detection technologies should be actively used to reduce the number of comparisons through LLM.

\begin{table}[]
\caption{Average Response Consistency Rate Range ($\omega$) Effected by \textit{temp} and Prompt}
\label{tab:RQ4:ConsistencyRateRange}

\resizebox{0.49\textwidth}{!}{
\renewcommand{\arraystretch}{0.3}

\begin{tabular}{c|c|c|c|c|c}
\toprule

 & \multicolumn{2}{c|}{ Effected by \textit{temp}} & & \multicolumn{2}{c}{Effected by prompt} \\ \hline
\multirow{6}{*}{Model}          & \multirow{6}{*}{Prompt} & \multirow{6}{*}{\begin{tabular}[c]{@{}c@{}} $\omega$\end{tabular} } & \multirow{6}{*}{Comp.}  & \multirow{6}{*}{\textit{temp}} & \multirow{6}{*}{\begin{tabular}[c]{@{}c@{}} $\omega$\end{tabular}} \\
                                &                         &                                            &                               &                              &                                            \\
                                &                         &                                            &                               &                              &                                            \\
                                &                         &                                            &                               &                              &                                            \\
                                &                         &                                            &                               &                              &                                            \\
                                &                         &                                            &                               &                              &                                            \\ \midrule

\multirow{12}{*}{\textit{GPT-4o}}        & \multirow{3}{*}{\textit{P0}}      & \multirow{3}{*}{0.032}                     & \multirow{12}{*}{$\approx$}   & \multirow{4}{*}{0}           & \multirow{4}{*}{0.025}                     \\
                                &                         &                                            &                               &                              &                                            \\
                                &                         &                                            &                               &                              &                                            \\ \cline{2-3}
                                & \multirow{3}{*}{\textit{P1}}      & \multirow{3}{*}{0.031}                     &                               &                              &                                            \\ \cline{5-6} 
                                &                         &                                            &                               & \multirow{4}{*}{0.3}         & \multirow{4}{*}{0.024}                     \\
                                &                         &                                            &                               &                              &                                            \\ \cline{2-3}
                                & \multirow{3}{*}{\textit{P2}}      & \multirow{3}{*}{0.078}                     &                               &                              &                                            \\
                                &                         &                                            &                               &                              &                                            \\ \cline{5-6} 
                                &                         &                                            &                               & \multirow{4}{*}{1}           & \multirow{4}{*}{0.060}                      \\ \cline{2-3}
                                & \multirow{3}{*}{\textit{P3}}      & \multirow{3}{*}{0.047}                     &                               &                              &                                            \\
                                &                         &                                            &                               &                              &                                            \\
                                &                         &                                            &                               &                              &                                            \\ \midrule
\multirow{12}{*}{\textit{GPT-4o-mini}}   & \multirow{3}{*}{\textit{P0}}      & \multirow{3}{*}{0.056}                     & \multirow{12}{*}{$\approx$}   & \multirow{4}{*}{0}           & \multirow{4}{*}{0.012}                     \\
                                &                         &                                            &                               &                              &                                            \\
                                &                         &                                            &                               &                              &                                            \\ \cline{2-3}
                                & \multirow{3}{*}{\textit{P1}}      & \multirow{3}{*}{0.035}                     &                               &                              &                                            \\ \cline{5-6} 
                                &                         &                                            &                               & \multirow{4}{*}{0.3}         & \multirow{4}{*}{0.003}                     \\
                                &                         &                                            &                               &                              &                                            \\ \cline{2-3}
                                & \multirow{3}{*}{\textit{P2}}      & \multirow{3}{*}{0.071}                     &                               &                              &                                            \\
                                &                         &                                            &                               &                              &                                            \\ \cline{5-6} 
                                &                         &                                            &                               & \multirow{4}{*}{1}           & \multirow{4}{*}{0.023}                     \\ \cline{2-3}
                                & \multirow{3}{*}{\textit{P3}}      & \multirow{3}{*}{0.059}                     &                               &                              &                                            \\
                                &                         &                                            &                               &                              &                                            \\
                                &                         &                                            &                               &                              &                                            \\ \midrule
\multirow{12}{*}{\textit{Llama 3.1}}      & \multirow{3}{*}{\textit{P0}}      & \multirow{3}{*}{0.003}                     & \multirow{12}{*}{\textless{}} & \multirow{4}{*}{0}           & \multirow{4}{*}{0.418}                     \\
                                &                         &                                            &                               &                              &                                            \\
                                &                         &                                            &                               &                              &                                            \\ \cline{2-3}
                                & \multirow{3}{*}{\textit{P1}}      & \multirow{3}{*}{0.010}                      &                               &                              &                                            \\ \cline{5-6} 
                                &                         &                                            &                               & \multirow{4}{*}{0.3}         & \multirow{4}{*}{0.415}                     \\
                                &                         &                                            &                               &                              &                                            \\ \cline{2-3}
                                & \multirow{3}{*}{\textit{P2}}      & \multirow{3}{*}{0.011}                     &                               &                              &                                            \\
                                &                         &                                            &                               &                              &                                            \\ \cline{5-6} 
                                &                         &                                            &                               & \multirow{4}{*}{2}           & \multirow{4}{*}{0.424}                     \\ \cline{2-3}
                                & \multirow{3}{*}{\textit{P3}}      & \multirow{3}{*}{0.298}                     &                               &                              &                                            \\
                                &                         &                                            &                               &                              &                                            \\
                                &                         &                                            &                               &                              &                                            \\ \midrule
\multirow{12}{*}{\textit{Mistral}}       & \multirow{3}{*}{\textit{P0}}      & \multirow{3}{*}{-0.003}                    & \multirow{12}{*}{\textless{}} & \multirow{4}{*}{0}           & \multirow{4}{*}{0.180}                      \\
                                &                         &                                            &                               &                              &                                            \\
                                &                         &                                            &                               &                              &                                            \\ \cline{2-3}
                                & \multirow{3}{*}{\textit{P1}}      & \multirow{3}{*}{0.020}                      &                               &                              &                                            \\ \cline{5-6} 
                                &                         &                                            &                               & \multirow{4}{*}{0.3}         & \multirow{4}{*}{0.185}                     \\
                                &                         &                                            &                               &                              &                                            \\ \cline{2-3}
                                & \multirow{3}{*}{\textit{P2}}      & \multirow{3}{*}{-0.007}                    &                               &                              &                                            \\
                                &                         &                                            &                               &                              &                                            \\ \cline{5-6} 
                                &                         &                                            &                               & \multirow{4}{*}{1}           & \multirow{4}{*}{0.198}                     \\ \cline{2-3}
                                & \multirow{3}{*}{\textit{P3}}      & \multirow{3}{*}{0.003}                     &                               &                              &                                            \\
                                &                         &                                            &                               &                              &                                            \\
                                &                         &                                            &                               &                              &                                            \\ \midrule 
                                
                                \multirow{3}{*}{\textit{o3-mini}} & \multirow{3}{*}{-} & \multirow{3}{*}{-} &\multirow{3}{*}{-} & \multirow{3}{*}{-} & \multirow{3}{*}{0.040} \\ &&&&&\\ &&&&& \\  \bottomrule

\end{tabular}}

\end{table}

\begin{table}[]
    \centering
    \captionsetup{justification=centering}

        \caption{Average F1 score Range of each Model \\  Lowest \textit{temp}: 0  ; Highest \textit{temp}: 2 for \textit{Llama 3.1} and 1 for the others.\noindent}
    \label{tab:averageF1Range}
\renewcommand{\arraystretch}{0.4}

    \resizebox{0.49\textwidth}{!}{

\begin{tabular}{cccc}
\toprule
\multirow{2}{*}{Model} & \multicolumn{3}{c}{Range of F1 score} \\ \cmidrule(lr){2-4} 
                       & \multicolumn{1}{c}{Lowest \textit{temp}}      & \multicolumn{1}{c}{0.3}   & Highest \textit{temp}   \\ \midrule
\textit{GPT-4o}                 & \multicolumn{1}{c}{0.012}   & \multicolumn{1}{c}{0.014}  & 0.012  \\ \midrule 
\textit{GPT-4o-mini}            & \multicolumn{1}{c}{0.009}   & \multicolumn{1}{c}{0.009}  & 0.021  \\ \midrule
\textit{Llama 3.1}               & \multicolumn{1}{c}{0.027}   & \multicolumn{1}{c}{0.026}  & 0.032  \\ \midrule
\textit{Mistral}                & \multicolumn{1}{c}{0.012}   & \multicolumn{1}{c}{0.015}  & 0.010 \\ \midrule \midrule
\textit{o3-mini}  & \multicolumn{3}{c}{0.012} \\ 
\bottomrule
\end{tabular}

}

\end{table}

\subsection{Threats to Validity}
\label{sec:threatsToValidity}
This study incorporated competitive programming data to construct datasets as the original research, where accepted code for the same problem is considered a clone. However, this ground truth may fail when using LLMs for code clone detection. For instance, competitive programming problems typically specify input data ranges and formats. For a problem with single-line input specifications, whether or not the submitted code handles multiple lines from the standard input does not affect the judgment. However, an LLM, unaware of such specifics, can classify code supporting multi-line input differently from code that does not, labeling them as non-clones. From the LLM's perspective, this judgment is accurate but would be counted as a false negative in evaluations. Therefore, the actual performance on \textit{CN}-related datasets could be slightly better than the observed results.

To automatically recognize detection results from LLMs' responses, current prompts instruct the LLM to output only ``yes'' or ``no'', ignoring the reasoning or explanations it would typically provide. This approach sacrifices information and does not fully use the capabilities of LLMs. For example, the issue mentioned in the previous paragraph cannot be addressed. However, automated result recognition remains a necessary function for code clone detectors. Improvements in this area will depend on further research advancements.

Fine-tuning has been widely demonstrated to considerably improve the performance of LLMs on specific tasks, including code clone detection. However, we chose not to fine-tune the models in this study for the following reasons. Fine-tuning LLMs using data from BigCloneBench would improve their detection accuracy on \textit{BCB-random} and \textit{BCB-average}. However, this study focused on evaluating whether the clone detection capabilities of LLMs can generalize to a broader range of datasets and, ultimately, to the entire codebase in the real world. Without establishing the representativeness of BigCloneBench for the entire coding domain or creating and evaluating models fine-tuned on datasets distinct from BigCloneBench and CodeNet to determine whether they maintain comparable accuracy on other datasets, the significance of achieving higher accuracy on BCB-related datasets would remain limited.

\section{Conclusion}
\label{sec:conclusions}
To investigate the generalization ability and response consistency of LLMs in code clone detection, a topic that has not been extensively explored in previous research, we constructed seven code clone datasets and then evaluated five LLMs on their detection performance. Our evaluation revealed two key findings. First, the LLMs demonstrated superior performance on datasets derived from \textit{CN} compared to \textit{BCB}, where their detection capabilities showed a notable decline. Second, most models exhibited high response consistency with minimal performance variations, suggesting that LLM-based code clone detection results are generally reproducible.
Future studies should focus on the following areas: exploring whether fine-tuning can enhance LLMs' detection performance across multiple datasets and whether LLMs can achieve similar performance for relatively less common programming languages.

\bibliographystyle{ACM-Reference-Format}


\end{document}